\documentclass[preprint]{revtex4-1}
\usepackage{graphicx}
\begin{document}
\preprint{}
\title{On growth of spinodal instabilities in nuclear matter }
\author{O. Yilmaz$^{1}$}\email{oyilmaz@metu.edu.tr}
\author{S. Ayik$^{2}$}\email{ayik@tntech.edu}
\author{F. Acar$^{1}$}
\author{A. Gokalp$^{3}$}

\affiliation{ $^{1}$Physics Department, Middle East Technical
University, 06800 Ankara, Turkey \\
$^{2}$Physics Department, Tennessee Technological University,
Cookeville, TN 38505, USA \\
$^{3}$ Department of Physics, Bilkent University, 06800 Ankara,
Turkey }
\date{\today}

\begin{abstract}
Early growth of density fluctuations of nuclear matter in spinodal region  is investigated employing the stochastic mean-field approach. In contrast to the earlier treatments in which only collective modes were included in the calculations, in the present work non-collective modes are also included, thus providing a complete treatment of the density correlation functions. Calculations are carried out for symmetric matter in non-relativistic framework using a semi-classical approximation.
\end{abstract}

\maketitle

\section{Introduction}

 Much work has been done to analyze multi-fragmentation processes observed in heavy-ion collisions. These processes exhibit many features compatible with liquid-gas phase transition of nuclear matter. Experimental measurements of charge correlations between produced fragments suggest that the induced phase transformation is due to the amplification of density fluctuations in the spinodal region. Microscopic description based on mean-field approximation can provide only a limited information about the spinodal dynamics. In the mean-field framework, it is only possible to determine the boundary of the dynamical instabilities and the growth rates of the most-unstable collective modes. Growth of density fluctuations and large amplitude dynamical evolution of the system require a microscopic framework beyond the mean-field that incorporates the dynamics of density fluctuations [1-5].

Considerable effort has been spent to improve the standard mean-field approximation by incorporating dynamics of density fluctuations [6-9]. In a recently proposed stochastic mean-field approach (SMF), the effect of quantal and thermal fluctuations in the initial state is incorporated into the description in a stochastic manner [10]. The standard mean-field approximation is a deterministic description in the sense that a well- defined initial state gives rise to a unique final state. On the other hand, in the SMF approach, starting from a proper distribution of initial states, specified by quantal and thermal fluctuations, an ensemble of mean-field trajectories are evolved each with its own self-consistent Hamiltonian. A number of recent publications have provided strong validation of the SMF approach for describing dynamics of density fluctuations at low energies [11-16].

 Employing the SMF approach in non-relativistic framework [17], as well as in relativistic framework based on Walecka-type effective field theory [18], we recently carried out a number of investigations of the early development of spinodal instabilities and baryon density correlation functions in nuclear matter in the  semi-classical and quantal descriptions [19-22]. In the linear response regime, linearizing the equation of motion around a suitable initial state inside the spinodal zone and employing the method of one-sided Fourier transform, it is possible to carry out nearly analytical treatment of the baryon density correlation functions. In these investigations only unstable collective modes were included in the calculations. As pointed out in Ref. [23], for a complete treatment of initial growth of density fluctuations it is necessary to include non-collective modes as well. In the present investigation, we undertake calculations of correlation functions of baryon density fluctuations including collective as well as non-collective modes.

 In Section 2, we present nearly analytical description of the density correlation functions, including collective and non-collective modes in the linear response framework, in the semi-classical limit. In Section 3, we present calculations at different initial densities and temperatures for symmetric matter. Conclusions are given in Section 4.

\section{Density correlation functions in spinodal region}

 The SMF approach goes beyond the standard mean-field theory by including quantal and thermal fluctuations in the initial state. An ensemble of single-particle density matrices are generated by incorporating the initial fluctuations.  It is possible to formulate the approach in relativistic and non-relativistic frameworks. Here we consider the non-relativistic approach. The single-particle density matrix of each event (indicated by the event label
$\lambda $), $\hat{\rho}^{\lambda} (\vec{r},\vec{r}',t)$, is evolved through to its own mean-field according to

 \begin{equation} \label{Eq1}
i\hbar \frac{\partial }{\partial t} \hat{\rho }_{a}^{\lambda } (t)=\left[h_{a}^{\lambda } (t),\hat{\rho }_{a}^{\lambda } (t)\right],
\end{equation}
where the label $a=p\uparrow ,p\downarrow ,n\uparrow ,n\downarrow $  indicate the isospin-spin quantum numbers, and $h_{a}^{\lambda } (t)$ is the mean-field Hamiltonian for the event. For the description of early growth of density fluctuations around an initial state $\hat{\rho }_{a} $, it is sufficient to linearize the equation of motion around this initial state. Subsequently small fluctuations, $\delta \hat{\rho }_{a}^{\lambda } (t)=\hat{\rho }_{a}^{\lambda } (t)-\hat{\rho }_{a} $, in the density matrix are determined by

\begin{equation} \label{Eq2}
i\hbar \frac{\partial }{\partial t} \delta \hat{\rho }_{a}^{\lambda } (t)=\left[h_{a} ,\delta \hat{\rho }_{a}^{\lambda } (t)\right]+\left[\delta U_{a}^{\lambda } (t),\hat{\rho }_{a} \right],
\end{equation}
where $h_{a}$ denotes the mean field Hamiltonian at the initial state and $\delta U_{a}^{\lambda } $ is the fluctuating part of the mean-field for the event. We are interested in investigating the development of spinodal instabilities in nuclear matter around a uniform initial state $\hat{\rho }_{a} $. In this case, by employing  a plane wave representation, we can carry out an almost analytical treatment of the density correlation function. In the plane wave basis, the linear response equations become

\begin{equation} \label{Eq3}
\begin{array}{l} {i\hbar \frac{\partial }{\partial t} <\vec{p}_{1} |\delta \hat{\rho }_{a}^{\lambda } (t)|\vec{p}_{2} >} \\ {=\left[\varepsilon _{a} (\vec{p}_{1} )-\varepsilon _{a} (\vec{p}_{2} )\right]<\vec{p}_{1} |\delta \hat{\rho }_{a}^{\lambda } (t)|\vec{p}_{2} >-\left[f_{a} (\vec{p}_{1} )-f_{a} (\vec{p}_{2} )\right]<\vec{p}_{1} |\delta U_{a}^{\lambda } (t)|\vec{p}_{2} >} , \end{array}
\end{equation}
here $f_{a} (\vec{p})$ denotes Fermi-Dirac factors at finite temperature. In the current work, we consider symmetric nuclear matter and assume that the mean-field potential depends only on the local nucleon density. Consequently, small fluctuations in the mean-field can be written as $\delta U_{a}^{\lambda } (\vec{r},t)=(\partial U/\partial \rho )_{0} \delta \rho _{}^{\lambda } (\vec{r},t)$, where $\delta \rho _{}^{\lambda } (\vec{r},t)$ denotes the small fluctuations in total nucleon density. Also, because of relative simplicity, we carry out calculations in the semi-classical limit. The  Fourier transform of nucleon density fluctuations, $\delta \tilde{\rho }_{}^{\lambda } (\vec{k},t)$ is related to the fluctuations of the density matrix according to

\begin{equation} \label{Eq4}
\delta \tilde{\rho }_{}^{\lambda } (\vec{k},t)=\sum _{a}\int \frac{d^{3} p}{(2\pi \hbar )^{3} }  <\vec{p}+\hbar \vec{k}/2|\delta \hat{\rho }_{a}^{\lambda } |\vec{p}-\hbar \vec{k}/2> ,
\end{equation}
where summation $a$ is over the spin-isospin quantum numbers. The method of one-sided Fourier transform is very useful for solving Eq. \ref{Eq3} [24,25]. We introduce the one-sided Fourier transform of the local density in time as

\begin{equation} \label{Eq5}
\delta \tilde{\rho }^{\lambda } (\vec{k},\omega )=\int _{0}^{\infty }dte^{i\omega t} \delta \tilde{\rho }^{\lambda}(\vec{k},t).
\end{equation}
This yield an algebraic equation for the Fourier transform of the local density, which can be solved to obtain

\begin{equation} \label{Eq6}
\delta \tilde{\rho }^{\lambda } (\vec{k},\omega )=-\frac{i}{\varepsilon (\vec{k},\omega )} G^{\lambda } (\vec{k},\omega ),
\end{equation}
where $\varepsilon (\vec{k},\omega )=1+F_{0} \chi (\vec{k},\omega )$ is the susceptibility, with $F_{0} =(\partial U/\partial \rho )_{0} $ as the zeroth order Landau parameter, and $\chi (\vec{k},\omega )$ denotes the Lindhard function. The semi-classical expression of the Lindhard function is given by

\begin{equation} \label{Eq7}
\chi (\vec{k},\omega )=-4\int \frac{d^{3} p}{(2\pi \hbar )^{3} }  \frac{\vec{v}\cdot \vec{k}}{\vec{v}\cdot \vec{k}-\omega } \frac{\partial f_{0} }{\partial \varepsilon } .
\end{equation}
The quantity $G^{\lambda } (\vec{k},\omega )$ in Eq.\ref{Eq6} is determined by the initial conditions
\begin{equation} \label{Eq8}
G^{\lambda } (\vec{k},\omega )=\sum _{a}\int \frac{d^{3} p}{(2\pi \hbar )^{3} }  \frac{<\vec{p}+\hbar \vec{k}/2|\delta \hat{\rho }_{a}^{\lambda } (0)|\vec{p}-\hbar \vec{k}/2>}{\vec{v}\cdot \vec{k}-\omega },
\end{equation}
which acts as a source for developing fluctuations. According to the basic postulate of the SMF approach, elements of the initial density matrix are uncorrelated Gaussian random numbers with zero mean values and with well defined variances. In the semi-classical limit their variances are given by
\begin{equation} \label{Eq9}
\begin{array}{l} {\overline{<\vec{p}+\hbar \vec{k}/2|\delta \hat{\rho }_{a}^{\lambda } (0)|\vec{p}-\hbar \vec{k}/2><\vec{p}'-\hbar \vec{k}'/2|\delta \hat{\rho }_{b}^{\lambda } (0)|\vec{p}'+\hbar \vec{k}'/2>}} \\ {=\delta _{ab} (2\pi \hbar )^{3} \delta (\vec{p}-\vec{p}')(2\pi)^{3}\delta (\vec{k}-\vec{k}')\left[f(\vec{p})\left(1-f(\vec{p})\right)\right]}. \end{array}
\end{equation}
Here and below the bar over the expressions indicates the ensemble average. If we ignore the Coulomb potential, this has the same form for neutrons and protons in symmetric nuclear matter, therefore we omit the label $a$ in the Fermi-Dirac factors. The factor $\delta _{ab} $ reflects the assumption that local density fluctuations in spin-isospin modes are uncorrelated in the initial state.

 According to the method of one-sided Fourier transform, time evolution of the density fluctuations is determined by the inverse transformation in time, which is expressed as a contour integral in the complex $\omega $-plane as

\begin{equation} \label{Eq10}
\delta \tilde{\rho }^{\lambda } (\vec{k},t)=-i\int _{-\infty +i\sigma }^{+\infty +i\sigma }\frac{d\omega }{2\pi }  \frac{G^{\lambda } (\vec{k},\omega )}{\varepsilon (\vec{k},\omega )} e^{-i\omega t},
\end{equation}
where integration path passes above all singularities of the integrand, as shown by line $C_{1} $ in Fig.1. We can calculate this integral by employing the residue theorem and closing the contour in a suitable manner. For the calculation of this integral, we need to investigate analytical structure of the integrand. In the spinodal zone, there are collective poles determined by imaginary solutions of the dispersion relation $\varepsilon (\vec{k},\omega )=0\to \omega =\pm i\Gamma _{k} $. The collective poles play an important role in early growth of density fluctuations. However, as pointed out by Bozek [23], collective poles alone do not give the full description of the growth of instabilities. In fact, collective pole contributions alone do not even satisfy the initial conditions.  By calculating the angular integral in the Lindhard function given by Eq.\ref{Eq7}, it is possible to see that, there is a cut singularity of the integrand in Eq.\ref{Eq10} along real axis in the complex $\omega$-plane. In addition, $G^{\lambda } (\vec{k},\omega )$ may also have singularities on the real axis. These non-collective poles of the integrand in Eq.\ref{Eq10} can be seen more clearly in discrete description of the plane wave representation.  If the system is placed in a finite box, plane waves are characterized by discrete values of momentum.  In this case as illustrated in [26], dispersion relation has both imaginary collective poles and non-collective poles on the real axis. In the continuous limit, these non-collective poles appear as a cut singularity. Since the integrand in complex $\omega $-plane is multivalued, the entire real $\omega $-axis is a branch cut. In order to calculate the integral in Eq.\ref{Eq10}, we choose the contour $C$, as shown in Fig. 1. We exclude the real $\omega $-axis by drawing the contour from $+\infty $ to the origin just above the real $\omega $-axis, and after jumping from the first Riemann surface to the second at the origin, we draw the contour just below the real $\omega $-axis from origin to $+\infty $ .  Contour is continued with a large semi-circle and it is completed by jumping from the second Riemann surface to the first one at origin,as shown in Fig.1. As a result, the integral can be expressed as

\begin{figure}[ht]
\hspace{1.0cm}
\includegraphics[width=15cm, height=21cm]{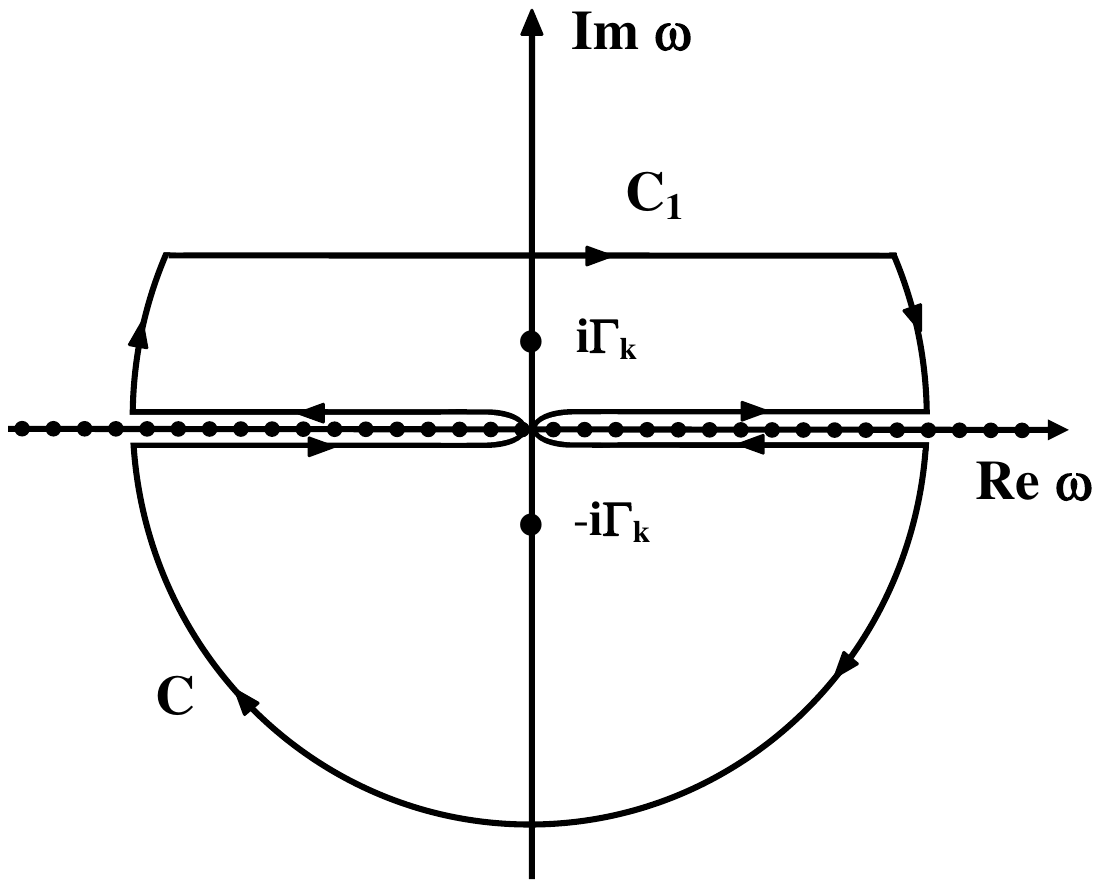}
\vspace{-14.0cm}
 \caption{\label{fig1}(Color online) The contour chosen to calculate the integral in Eq. \ref{Eq10}. Dots $\pm i\Gamma {}_{k} $ indicate the collective poles of the integrand, and the dotted line along the entire real $\omega $-axis is the branch cut of the integrand. }
\vspace{0.0cm}
\end{figure}

\begin{equation} \label{Eq11}
\delta \tilde{\rho }^{\lambda } (\vec{k},t)=\delta \tilde{\rho }_{pole}^{\lambda } (\vec{k},t)+\delta \tilde{\rho }_{cut}^{\lambda } (\vec{k},t),
\end{equation}
where the first term is the pole contribution
\begin{equation} \label{Eq12}
\delta \tilde{\rho }_{pole}^{\lambda } (\vec{k},t)=-\sum _{\pm }\frac{G^{\lambda } (\vec{k},\pm i\Gamma _{k} )}{\partial \varepsilon (\vec{k},\omega )/\partial \omega |_{\omega =\pm i\Gamma _{k} } }  e^{\pm \Gamma _{k} t},
\end{equation}
and the cut contribution is given by
\begin{equation} \label{Eq13}
\delta \tilde{\rho }_{cut}^{\lambda } (\vec{k},t)=-i\int _{-\infty }^{+\infty }\frac{d\omega }{2\pi } \left[\frac{G^{\lambda } (\vec{k},\omega +i\varepsilon )}{\varepsilon (\vec{k},\omega +i\varepsilon )} -\frac{G^{\lambda } (\vec{k},\omega -i\varepsilon )}{\varepsilon (\vec{k},\omega -i\varepsilon )} \right] e^{-i\omega t} .
\end{equation}
Local nucleon density fluctuations $\delta \rho ^{\lambda } (\vec{r}',t)$ are determined by the Fourier transform of $\delta \tilde{\rho }^{\lambda } (\vec{k},t)$ as
\begin{equation} \label{Eq14}
\delta \rho ^{\lambda } (\vec{r},t)=\int \frac{d^{3} k}{(2\pi )^{3} }  e^{i\vec{k}\cdot \vec{r}} \delta \tilde{\rho }^{\lambda } (\vec{k},t).
\end{equation}

In order to investigate the development of the spinodal instabilities a very useful quantity is the equal time correlation function of local density fluctuations

\begin{equation} \label{Eq15}
\sigma (|\vec{r}-\vec{r}'|,t)=\overline{\delta \rho ^{\lambda } (\vec{r},t)\delta \rho ^{\lambda } (\vec{r}',t)}=\int \frac{d^{3} k}{(2\pi )^{3} }  e^{i\vec{k}\cdot (\vec{r}-\vec{r}')} \sigma (\vec{k},t).
\end{equation}
The spectral intensity $\sigma (\vec{k},t)$ of the correlation function is determined in terms of the variance of the Fourier transform of density fluctuations according to

\begin{equation} \label{Eq16}
\sigma (\vec{k},t)(2\pi )^{3} \delta (\vec{k}-\vec{k}')=\overline{\delta \tilde{\rho }^{\lambda } (\vec{k},t)\delta \tilde{\rho }^{\lambda } (-\vec{k}',t)}  .
\end{equation}
We can evaluate the spectral intensity $\sigma (\vec{k},t)$ by evaluating the ensemble average using the Eqs. (\ref{Eq12}) and (\ref{Eq13})
for pole $\delta \tilde{\rho }_{pole}^{\lambda } (\vec{k},t)$ and cut $\delta \tilde{\rho }_{cut}^{\lambda } (\vec{k},t)$ part of the Fourier transform of density fluctuations and the Eq. (\ref{Eq9}) for the initial fluctuations. As a result, the spectral intensity becomes

\begin{equation} \label{Eq17}
\sigma (\vec{k},t)=\sigma_{pp} (\vec{k},t)+2\sigma_{pc} (\vec{k},t)+\sigma_{cc} (\vec{k},t),
\end{equation}
where the first and last terms are due to pole and cut parts of the spectral intensity and the middle term denotes the cross contribution. The pole part is
\begin{equation} \label{Eq18}
\sigma_{pp} (\vec{k},t)=\frac{E_{+} }{|\left[\partial \varepsilon (\vec{k},\omega )/\partial \omega \right]_{\omega =i\Gamma _{k} } |^{2} } \left(e^{+2\Gamma _{k} t} +e^{-2\Gamma _{k} t} \right)-\frac{2E_{-} }{|\left[\partial \varepsilon (\vec{k},\omega )/\partial \omega \right]_{\omega =i\Gamma _{k} } |^{2} },
\end{equation}
where
\begin{equation} \label{Eq19}
E_{\pm } =4\int \frac{d^{3} p}{(2\pi \hbar )^{3} } f(\vec{p})\left(1-f(\vec{p})\right) \frac{(\vec{v}\cdot \vec{k})^{2} \pm \Gamma _{k}^{2} }{\left[(\vec{v}\cdot \vec{k})^{2} +\Gamma _{k}^{2} \right]^{2} }.
\end{equation}
The pole contributions are the same as what we found in our previous investigation [17]. The cut-cut contribution has four terms

\begin{equation} \label{Eq20}
\sigma_{cc} (\vec{k},t)=A_{\mp } (\vec{k},t)+\tilde{A}_{\mp } (\vec{k},t)+\tilde{A}_{\pm } (\vec{k},t)+A_{\pm } (\vec{k},t),
\end{equation}
with
\begin{equation} \label{Eq21}
A_{\mp } (\vec{k},t)=\int _{-\infty }^{+\infty }\frac{d\omega }{2\pi }  \int _{-\infty }^{+\infty }\frac{d\omega '}{2\pi }  \frac{\phi (\omega \mp i\eta )+\phi (\omega '\mp i\eta )}{\varepsilon (\vec{k},\omega \mp i\eta )\varepsilon (\vec{k},\omega '\mp i\eta )} \frac{1}{\omega +\omega '\mp 2i\eta } e^{-i(\omega +\omega ')t},
\end{equation}
and
\begin{equation} \label{Eq22}
\tilde{A}_{\mp } (\vec{k},t)=-\int _{-\infty }^{+\infty }\frac{d\omega }{2\pi }  \int _{-\infty }^{+\infty }\frac{d\omega '}{2\pi }  \frac{\phi (\omega \mp i\eta )+\phi (\omega '\pm i\eta )}{\varepsilon (\vec{k},\omega \mp i\eta )\varepsilon (\vec{k},\omega '\pm i\eta )} \frac{1}{\omega +\omega '} e^{-i(\omega +\omega ')t} .
\end{equation}
In these expressions $\eta $ is an infinitesimal positive number, and quantity $\phi (\omega +i\eta )$ is defined by
\begin{equation} \label{Eq23}
\phi (\omega +i\eta )=4\int \frac{d^{3} p}{(2\pi \hbar )^{3} }  f(\vec{p})\left(1-f(\vec{p})\right)\frac{1}{\vec{v}\cdot \vec{k}-\omega -i\eta } .
\end{equation}
In the double integral of $A_{\mp } (\vec{k},t)$, there are principle value and delta function contributions which are identified using the identity
$1/\left(\omega +\omega '\mp i\eta \right)=P(1/\omega +\omega ')\pm i\eta \delta (\omega +\omega ')$. The integrand of $\tilde{A}_{\mp } (\vec{k},t)$, in contrast to its appearance, is a well behaved  function, because when $\omega '=-\omega $, the nominator is also zero therefore the ratio
$\left(\phi (\omega \mp i\eta )+\phi (\omega '\pm i\eta )\right)/\left(\omega +\omega '\right)$ becomes finite.

Mixed term in the spectral intensity of Eq. (\ref{Eq17}) has the following form

\begin{equation} \label{Eq24}
\sigma_{pc} (\vec{k},t)=B_{\mp } (\vec{k},t)+\tilde{B}_{\mp } (\vec{k},t)+\tilde{B}_{\pm } (\vec{k},t)+B_{\pm } (\vec{k},t),
\end{equation}
with
\begin{equation} \label{Eq25}
B_{\mp } (\vec{k},t)=\pm i\frac{e^{\mp \Gamma _{k} t} }{\partial \varepsilon (\vec{k},\omega )/\partial \omega |_{\omega =\mp i\Gamma _{k} } } \int _{-\infty }^{+\infty }\frac{d\omega }{2\pi }  \frac{\phi (\mp i\Gamma _{k} )+\phi (\omega \mp i\eta )}{\varepsilon (\vec{k},\omega \mp i\eta )} \frac{1}{\omega \mp i\Gamma _{k} } e^{-i\omega t},
\end{equation}
and
\begin{equation} \label{Eq26}
\tilde{B}_{\mp } (\vec{k},t)=\mp i\frac{e^{\mp \Gamma _{k} t} }{\partial \varepsilon (\vec{k},\omega )/\partial \omega |_{\omega =\mp i\Gamma _{k} } } \int _{-\infty }^{+\infty }\frac{d\omega }{2\pi }  \frac{\phi (\mp i\Gamma _{k} )+\phi (\omega \pm i\eta )}{\varepsilon (\vec{k},\omega \pm i\eta )} \frac{1}{\omega \mp i\Gamma _{k} } e^{-i\omega t} .
\end{equation}

\begin{figure}[b]
\hspace{-0.5cm}
\includegraphics[width=14cm, height=9cm]{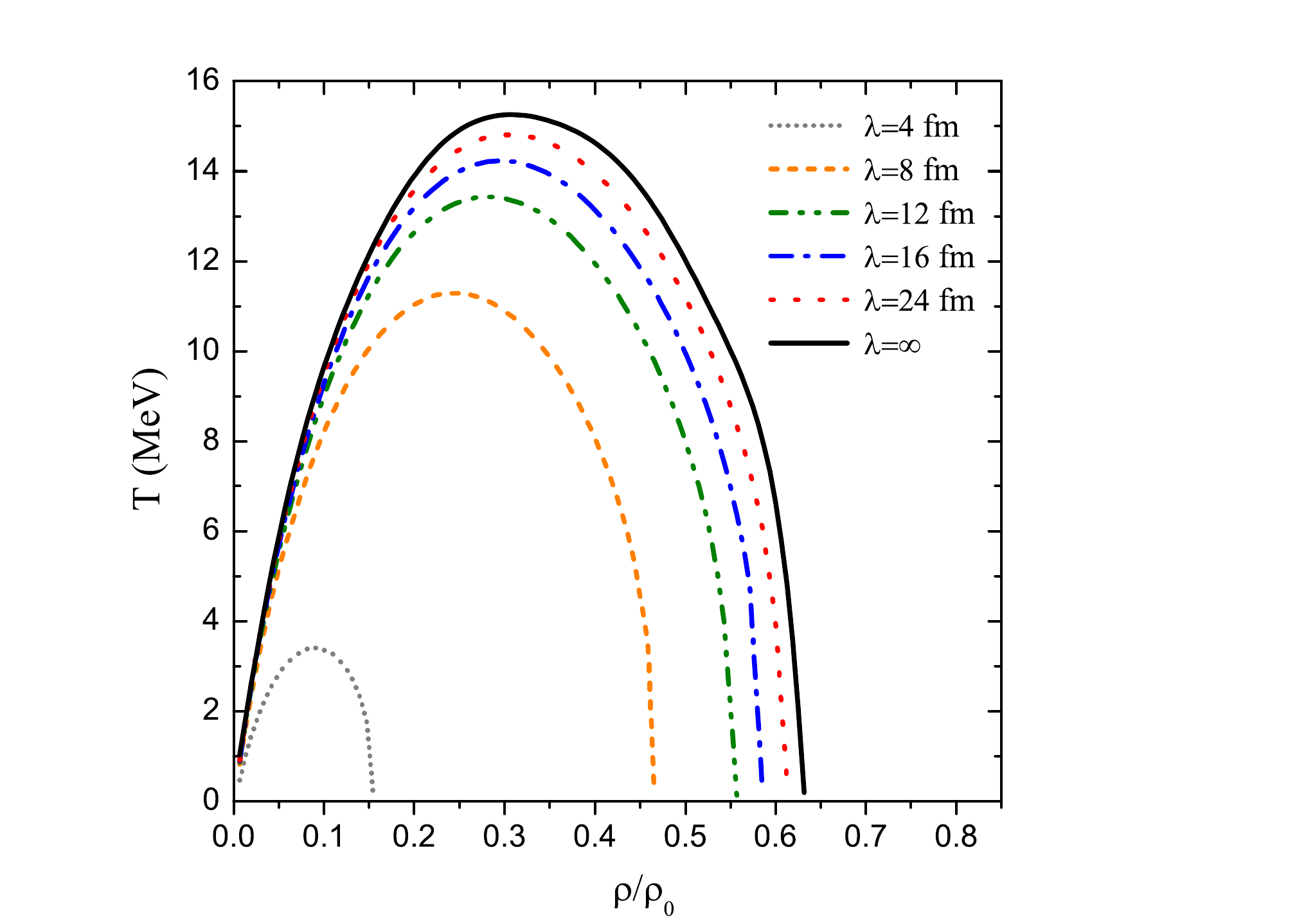}
\vspace{-0.5cm}
 \caption{\label{fig2}(Color online)Phase diagram in density-temperature plane for different wavelengths corresponding to the potential given by
Eq. (\ref{Eq27}). }
\vspace{-0.0cm}
\end{figure}

\section{Results}

In the numerical calculations we employ an effective Skyrme potential given in Ref. [6]. For symmetric nuclear matter it has the form
\begin{equation} \label{Eq27}
U(\rho )=A\left(\frac{\rho }{\rho _{0} } \right)+B\left(\frac{\rho }{\rho _{0} } \right)^{\alpha +1} -D\nabla ^{2} \rho,
\end{equation}
where $\rho (\vec{r},t)$ is the local nucleon density. The numerical parameters $A=-356.8MeV$, $B=+303.9MeV$, $\alpha =1/6$, and $D=+130.0MeV\cdot fm^{5} $ are adjusted to reproduce saturation properties of symmetric nuclear matter:  binding energy $\varepsilon _{0} =15.7~MeV$/nucleon, zero pressure at saturation density $\rho _{0} =0.16fm^{-3} $, compressibility modulus $K=201.0MeV$, and the surface energy coefficient of the Weizsacker mass formula $a_{surf} =18.6MeV$. Fig. 2 shows phase diagrams corresponding to different wavelengths in temperature-density plane. These diagrams indicate the boundaries of spinodal unstable regions for different wavelengths, starting from upper boundary for $\lambda =\infty $.  The critical temperature corresponding to this effective Skyrme potential is $T_{c} =15.5MeV$, which occurs approximately at a density $\rho =0.3\rho _{0} $.

\begin{figure}[h]
\hspace{-0.5cm}
\includegraphics[width=11cm, height=13cm]{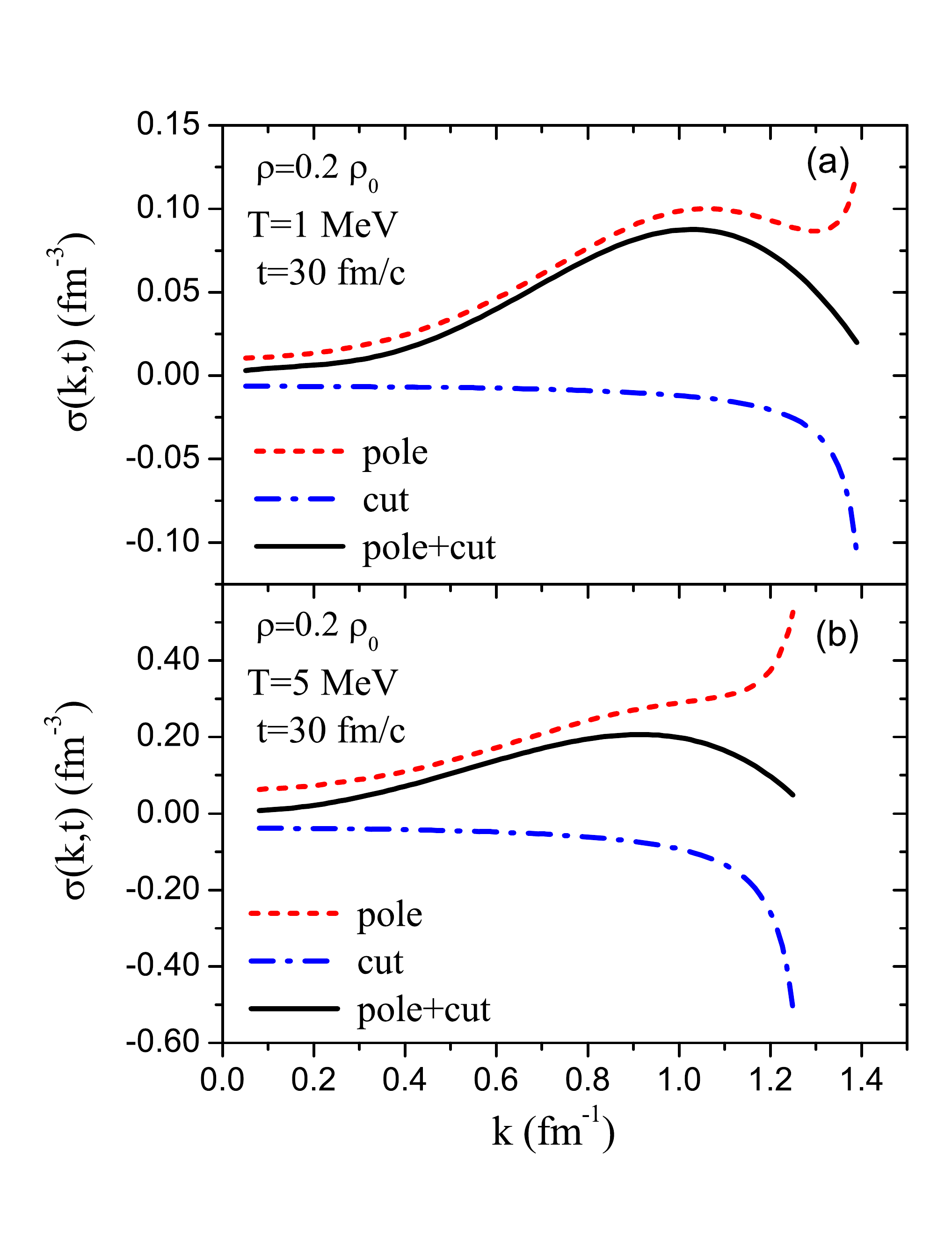}
\vspace{-1.5cm}
 \caption{\label{fig3}(Color online) Spectral intensity of the correlation function as a function of wave number at initial density $\rho =0.2\rho _{0} fm^{-3} $ at time $t=30fm/c$ at temperature $T=1MeV$ (a) and $T=5MeV$ (b). Dotted, dashed-dotted and solid lines are results of pole, cut, and total contributions, respectively. }
\vspace{-0.0cm}
\end{figure}

\begin{figure}[h]
\hspace{-0.5cm}
\includegraphics[width=11cm, height=13cm]{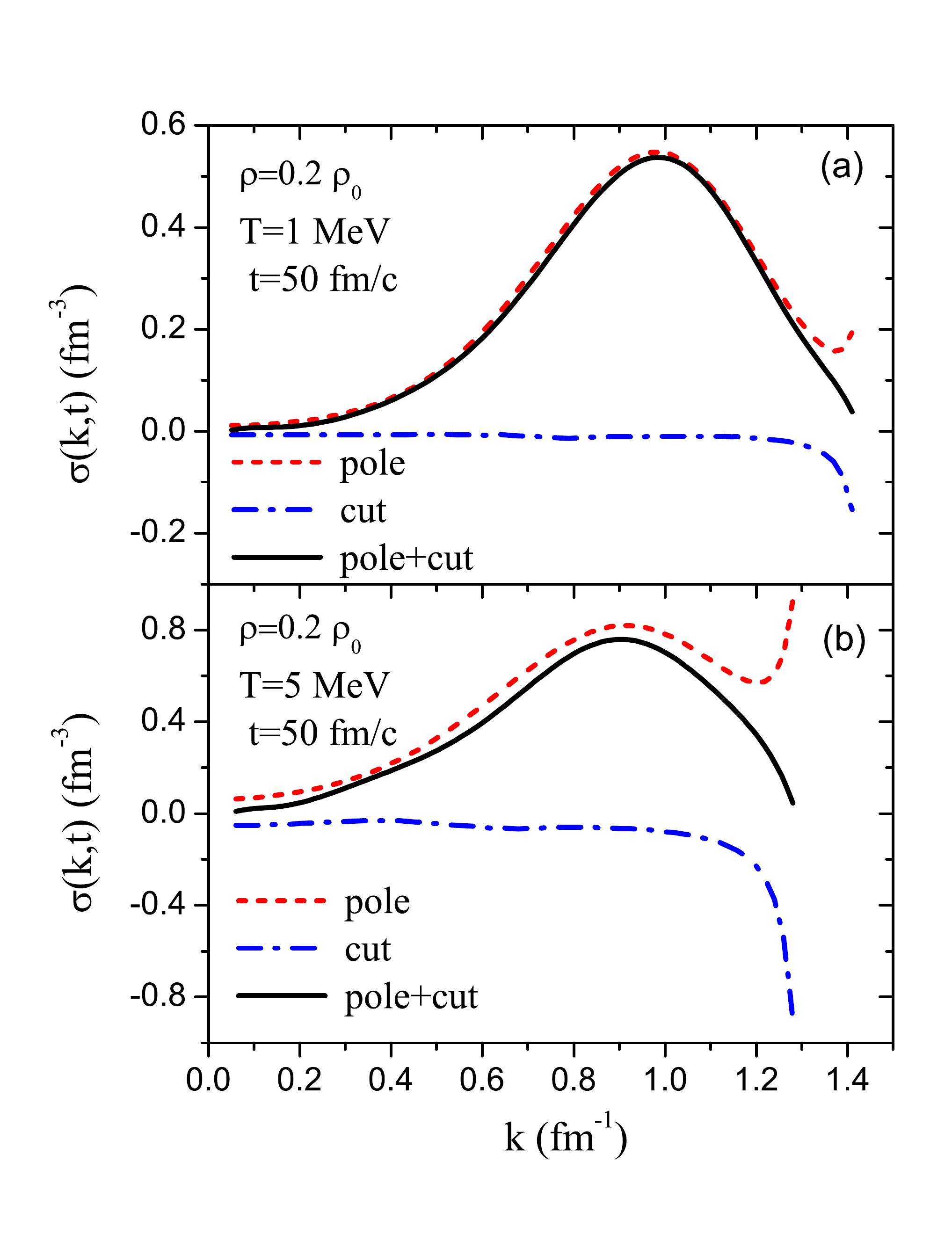}
\vspace{-1.5cm}
 \caption{\label{fig4}(Color online) Spectral intensity of the correlation function as a function of wave number at initial density $\rho =0.2\rho _{0} fm^{-3} $ at time $t=50fm/c$ at temperature $T=1MeV$ (a) and $T=5MeV$ (b). Dotted, dashed-dotted and solid lines are results of pole, cut, and total contributions, respectively.}
\vspace{-0.0cm}
\end{figure}

\begin{figure}[h]
\hspace{-0.5cm}
\includegraphics[width=11cm, height=13cm]{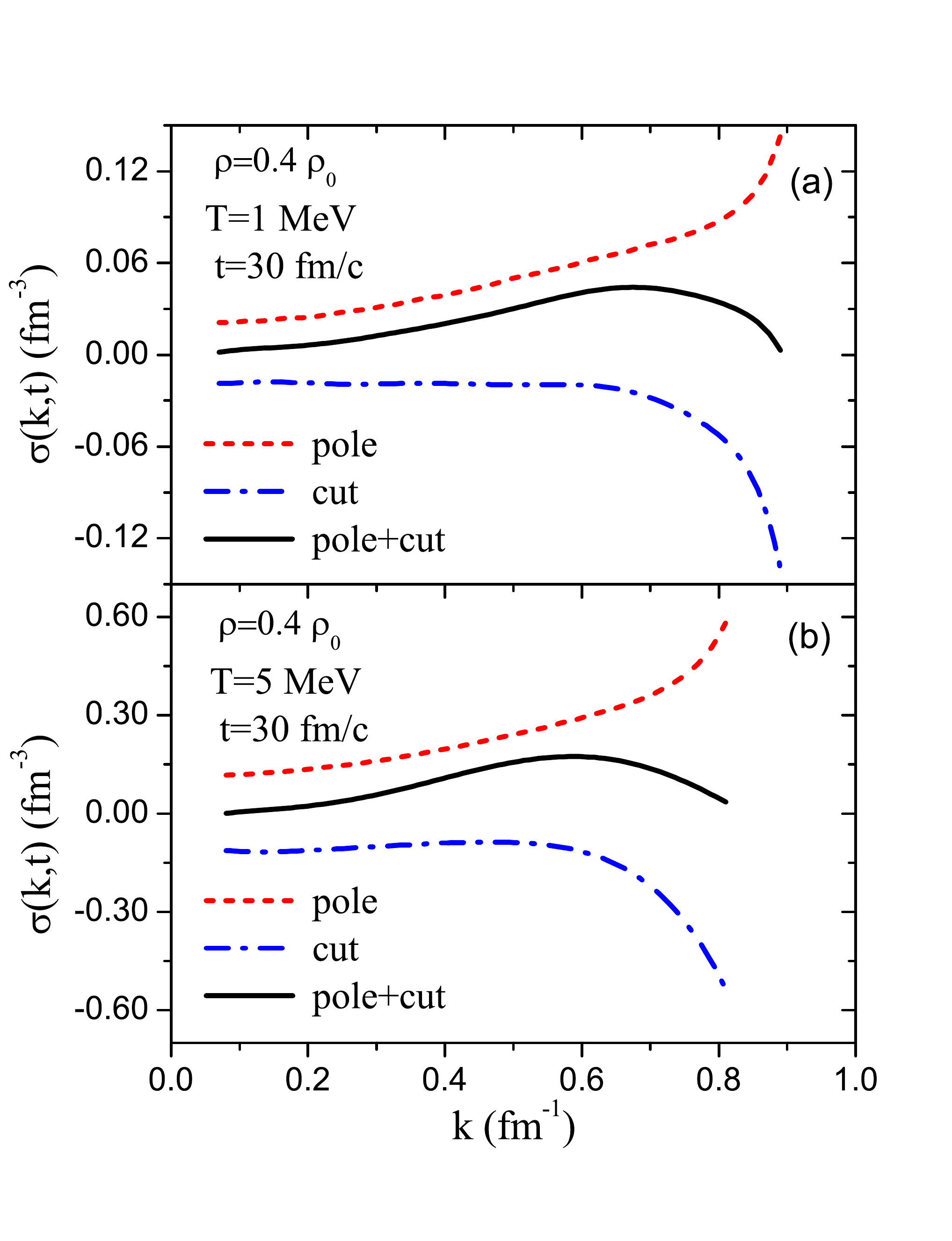}
\vspace{-1.5cm}
 \caption{\label{fig5}(Color online) Spectral intensity of the correlation function as a function of wave number at initial density $\rho =0.4\rho _{0} fm^{-3} $ at time $t=30fm/c$ at temperature $T=1MeV$ (a) and $T=5MeV$ (b). Dotted, dashed-dotted and solid lines are results of pole, cut, and total contributions, respectively. }
\vspace{-0.0cm}
\end{figure}

\begin{figure}[h]
\hspace{-0.5cm}
\includegraphics[width=11cm, height=13cm]{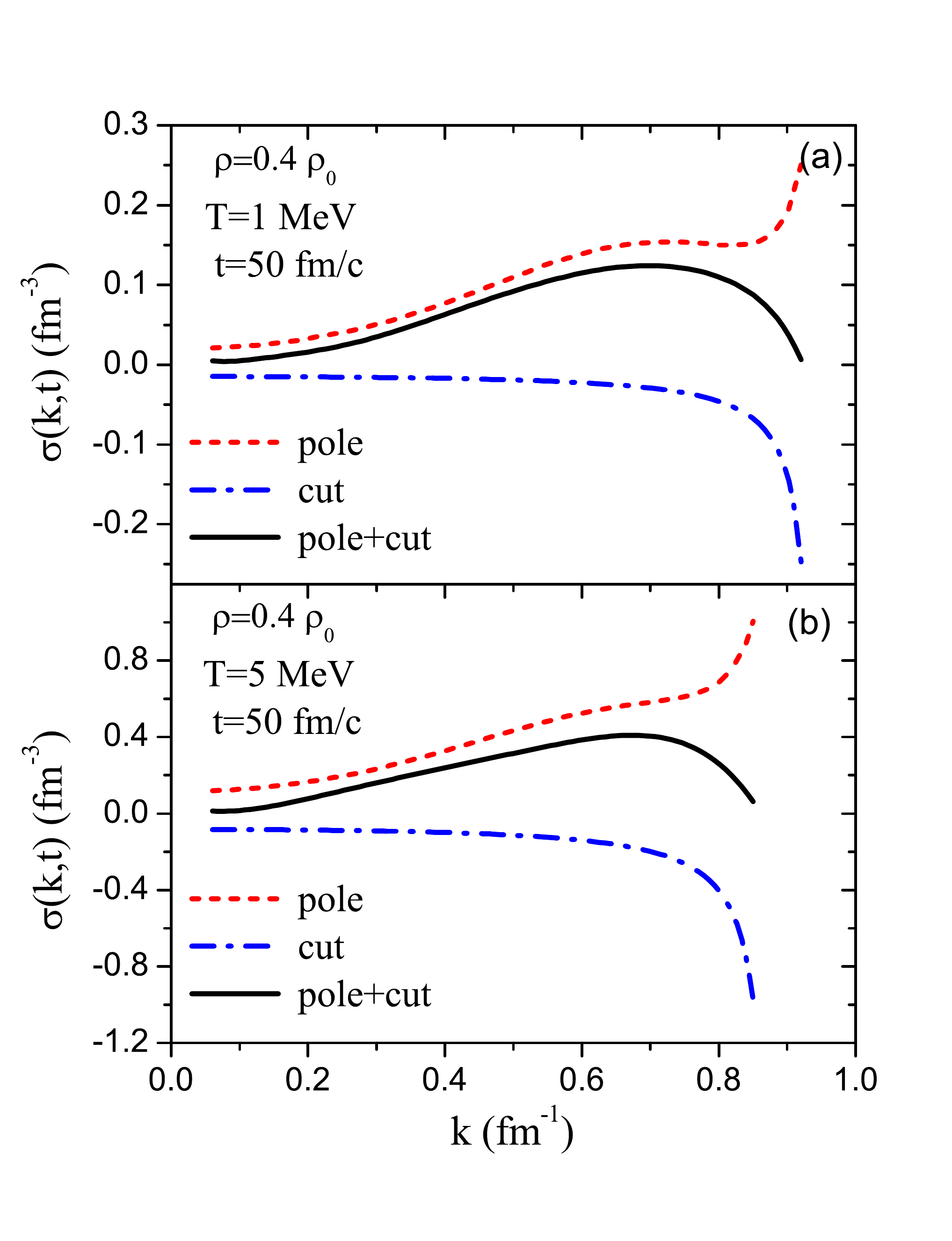}
\vspace{-1.5cm}
 \caption{\label{fig6}(Color online) Spectral intensity of the correlation function as a function of wave number at initial density $\rho =0.4\rho _{0} fm^{-3} $ at time $t=50fm/c$ at temperature $T=1MeV$(a) and $T=5MeV$(b). Dotted, dashed-dotted and solid lines are results of pole, cut, and total contributions, respectively.}
\vspace{-0.0cm}
\end{figure}

We pick two reference states with initial densities $\rho =0.2\rho _{0} $ and $\rho =0.4\rho _{0} $, and calculate equal time correlation function of density fluctuations at two different temperatures $T=1.0MeV$and $T=5.0MeV$.  In Figs.(3-6), we plot spectral intensity $\sigma(\vec{k},t)$ of correlation functions as a function of wave number $k$ at two different times $t=30fm/c$ and $t=50fm/c$ for two different temperatures and  initial densities as indicated above. At each initial density and temperature, the upper limit of the wave number $k_{\max } $ is determined by the condition that the inverse growth rate of the mode vanishes,
$\Gamma _{k} =0$.  Dashed, dash-dotted and solid lines indicate the result of calculations of Eq. (\ref{Eq17}) with pole contributions only, with cut contributions only, and the total of all terms. The cut contributions include cut-cut contribution $\sigma{}_{cc} (\vec{k},t)$, and the cross terms due to pole and cut parts
$2\sigma_{pc} (\vec{k},t)$. From these figures we make two important observations: Cut terms make an important negative contribution during the early phase of growth, hence slowing down the growth of instabilities. During later times, collective poles dominate the growth of density fluctuations, and the cut terms representing the effects of non-collective poles do not grow in time, as discussed in earlier studies [1]. We also observe that both pole and cut contributions have divergent behavior with opposite signs, as wave numbers approach their upper limit, $k\to k{}_{\max } $. These divergent behaviors cancel each other out to produce a nice regular behavior of the spectral intensity as a function of wave number.

\begin{figure}[h]
\hspace{-0.5cm}
\includegraphics[width=11cm, height=13cm]{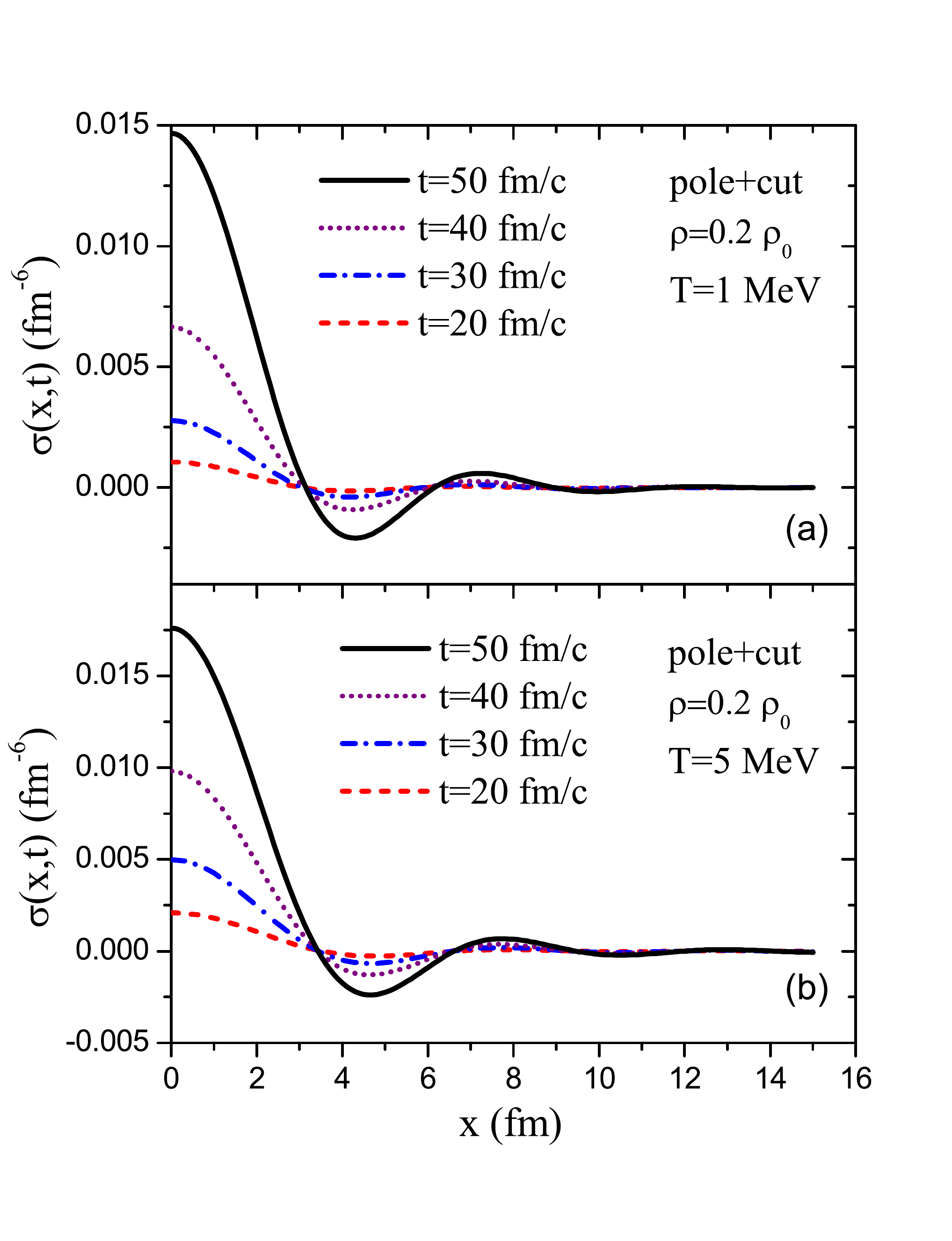}
\vspace{-1.5cm}
 \caption{\label{fig7}(Color online) Density correlation function as a function of distance between two space location $x=|\vec{r}-\vec{r}¨'~|$ for initial density $\rho =0.2\rho _{0} fm^{-3} $ and temperature $T=1MeV$(a) and $T=5MeV$(b) at times $t=20,30,40,50fm/c$.}
\vspace{-0.0cm}
\end{figure}

\begin{figure}[h]
\hspace{-0.5cm}
\includegraphics[width=11cm, height=13cm]{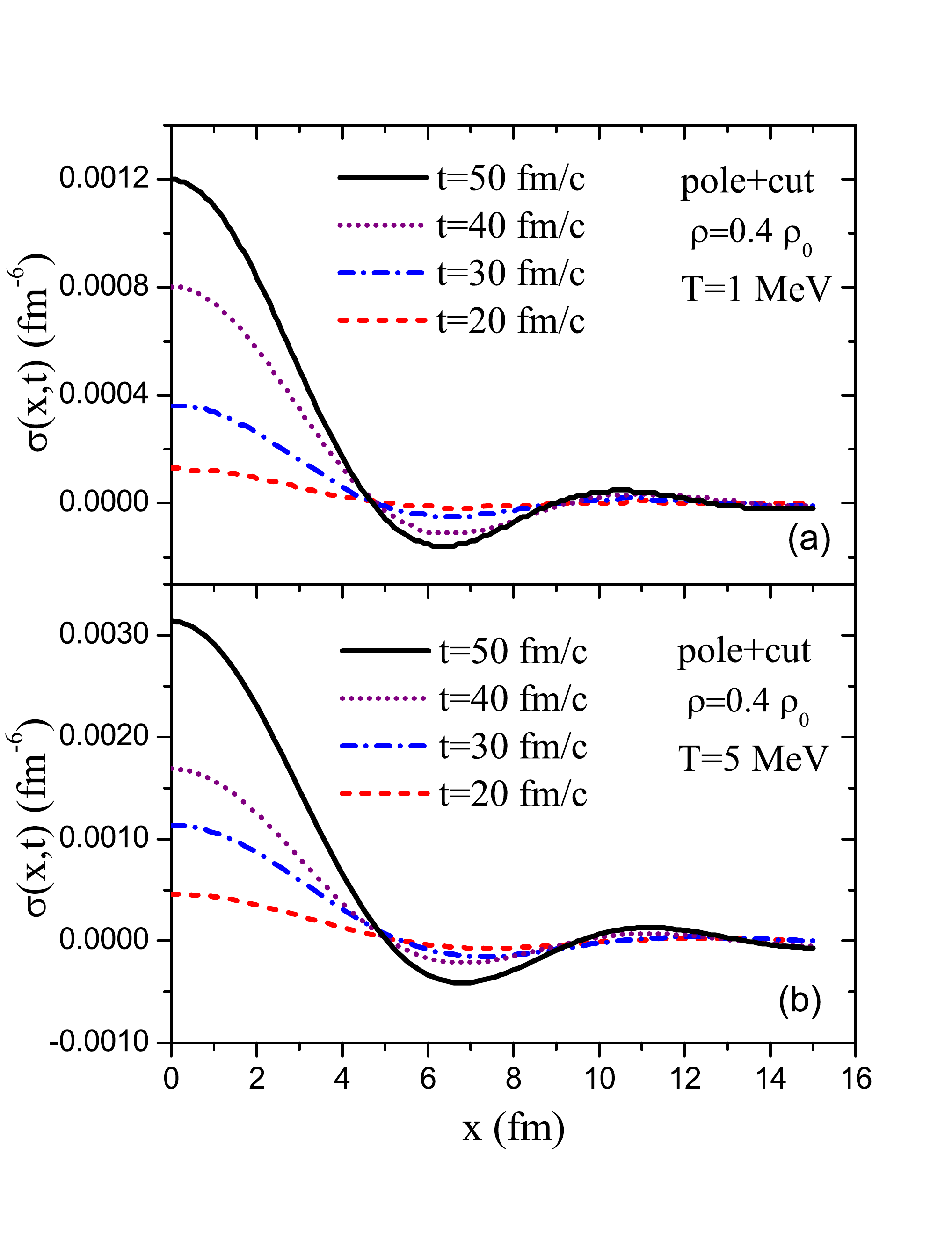}
\vspace{-1.5cm}
 \caption{\label{fig8}(Color online) Density correlation function as a function of distance between two space location $x=|\vec{r}-\vec{r}~'|$ for initial density$\rho =0.4\rho _{0} fm^{-3} $and temperature $T=1MeV$(a) and $T=5MeV$(b) at times $t=20,30,40,50fm/c$. }
\vspace{-0.0cm}
\end{figure}

Figs. (7-8) show the equal time correlation functions $\sigma (|\vec{r}-\vec{r}~'|,t)$ of density fluctuations as a function of distance between two space locations $x=|\vec{r}-\vec{r}~'|$ at two different initial densities and temperatures corresponding to Figs.(3-6). The density correlation functions calculated with full spectral intensities are shown by solid lines in Figs. (3-6), and the results are plotted in Figs.(7-8)  at four different times $t=20,30,40,50fm/c$. Growth of density fluctuations is clearly visible in these figures. In addition, we can obtain information about typical size of condensation regions during the growth of fluctuations. For this purpose, we introduce correlation length $x_{C} $ as the width of the correlation function at half maximum. The correlation length provides a measure for the size of condensing droplets during the growth of fluctuations. In the correlation volume $\Delta V_{C} =4\pi x_{C}^{3} /3$, the variance of local density fluctuations at time $t$ is approximately given by $\sigma (x_{C} ,t)$. The number of nucleons in each correlation volume fluctuates with dispersions $\Delta A_{C} =\Delta V_{C} \sqrt{\sigma (x_{C} ,t)}$. As a result, the nucleon number in each correlation volume fluctuates approximately within the range $\Delta A_{0} -\Delta A_{C} \le \Delta A\le \Delta A_{0} +\Delta A_{C} $, where $\Delta A_{0} =\Delta V_{C} \rho $ denotes the number of nucleons at the initial uniform state. From this analysis we notice that the spinodal decomposition mechanism does not necessarily lead to similar size fragmentation of the system. Rather this mechanism leads to a mixture of various cluster sizes . As seen from Figs. (7-8), correlation length is not very sensitive to the temperature and time evolution, but depends on the initial density. The correlation lengths are $x_{C} =2.0fm$ and $x_{C} =3.0fm$ for the initial densities $\rho =0.2\rho _{0} fm^{-3} $ and $\rho =0.4\rho _{0} fm^{-3} $, respectively. As an example, we note that in the lower panel of Fig.\ref{Eq8},  at the initial density $\rho =0.4\rho _{0} fm^{-3} $and temperature $T=5.0MeV$,the magnitude of dispersion of density fluctuations at time $t=50fm/c$ is about $\sqrt{\sigma (x_{C} ,t)} =0.04fm^{-3} $. Consequently, the number of nucleons in the correlation volume approximately fluctuates in the range  $3\le \Delta A\le 12$.

We like to clarify that the linear response treatment provides a description for the early evolution of the condensation mechanism inside the spinodal region.  Since the density fluctuations keep growing nearly exponentially in time, the linear response treatment of fluctuations beyond a limiting time does not provide a realistic description of instabilities.  We can constraint the linear response regime approximately by the condition that the dispersion on nucleon density fluctuations remains below the average nucleon density in the correlation volume, i.e., $ \sqrt{\sigma (x_{C} ,t)}\le \rho $.  As seen in Fig. (8), this condition is well satisfied for the initial density $\rho =0.4\rho _{0} fm^{-3} $ for times up to $t=50 fm/c $. In a similar manner, we can estimate nucleon number fluctuations in the correlation volume for the initial density $\rho =0.2\rho _{0} fm^{-3} $. However, we note that according to the constraint,  as seen in  Fig. (7), at this low initial density the linear response regime is already exceeded beyond time $ t=25 fm/c $. We also note that, in the linear response treatment, even though density fluctuations and hence early size of condensing fragments grow in the correlation volume, the matter is not expanding, and hence size of the correlation volume remains nearly constant in time. In the linear response description, we can approximately specify the range of early cluster size distribution, however  we cannot identify a favored cluster size. For more realistic treatment of fragmentation mechanism, we need to study nonlinear evolution of the density fluctuations in the framework of the SMF approach. In such a non-linear evolution, it is expected that the exponential growth of correlation function of density fluctuations saturates. Furthermore, since in such a non-linear evolution the matter is expanding,  the size of correlation volume become enlarged in time and the fragment size distribution is modified by recombination and breakup of clusters produced during the early phase of the evolution.  But this study is beyond the description of the present work.

In previous studies [17,19-22], in the calculations of the density correlation functions we included only the contributions due to the collective pole in the residue integral in Eq. (\ref{Eq10}). As seen from dashed lines in Figs.(3-6),  we encounter the problem that both pole and cut contributions have divergent behavior as wave numbers approach its upper limit, $k\to k{}_{\max } $. To cure this problem in those calculations, we introduced a cut-off in the integral over the wave numbers in Eq. (\ref{Eq18}). At initial density of $\rho =0.2\rho _{0} fm^{-3} $ the cut-off value was taken at the local minimum point of the spectral intensity.
\begin{figure}[h]
\hspace{-0.5cm}
\includegraphics[width=11cm, height=13cm]{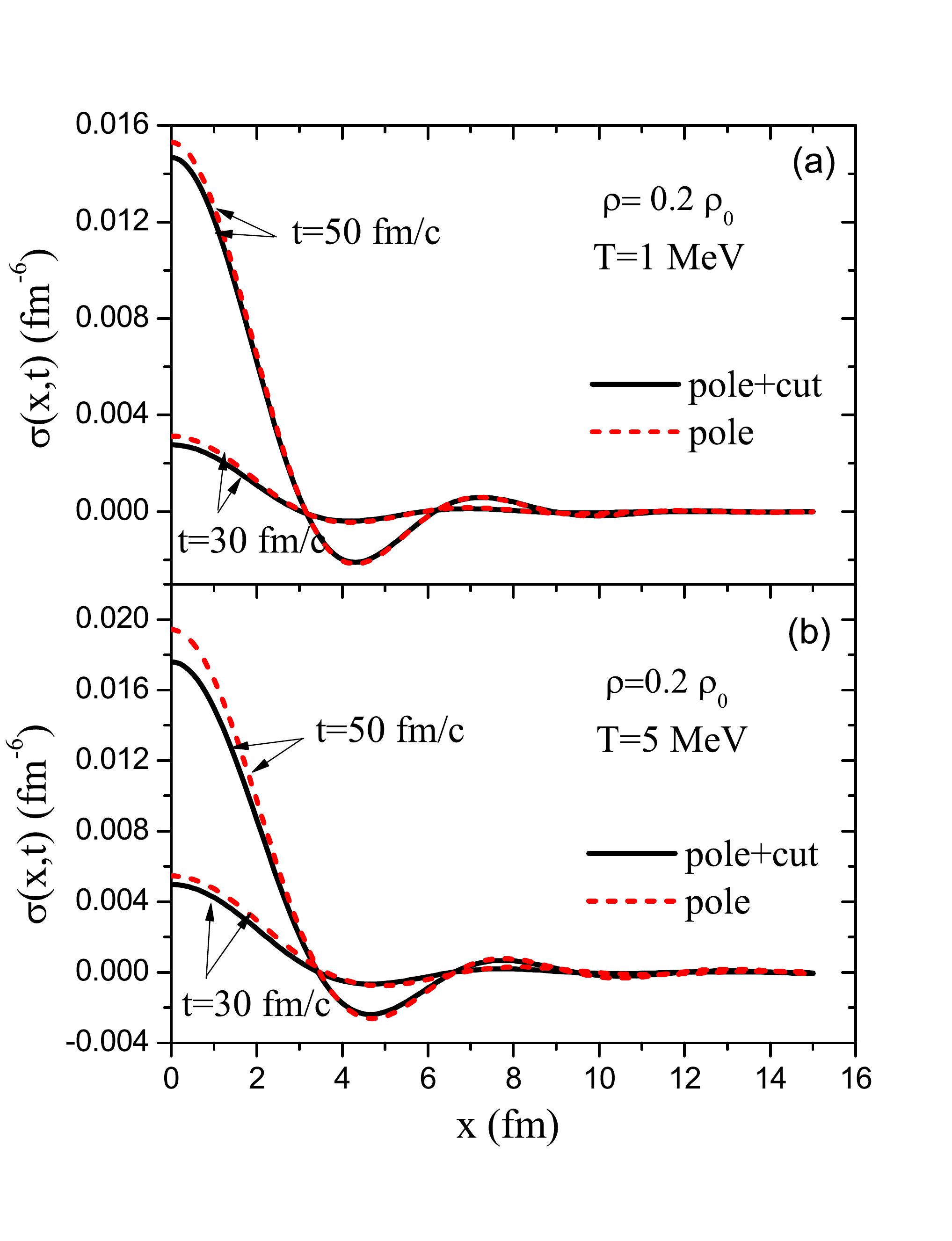}
\vspace{-1.5cm}
 \caption{\label{fig9}(Color online) Density correlation function as a function of distance between two space location $x=|\vec{r}-\vec{r}~'|$ for initial density$\rho =0.2\rho _{0} fm^{-3} $ and temperature $T=1MeV$ (a) and $T=5MeV$(b) at times $t=30,50fm/c$. Solid and dotted lines are complete calculations and pole contributions, respectively. }
\vspace{-0.0cm}
\end{figure}

\begin{figure}[h]
\hspace{-0.5cm}
\includegraphics[width=11cm, height=13cm]{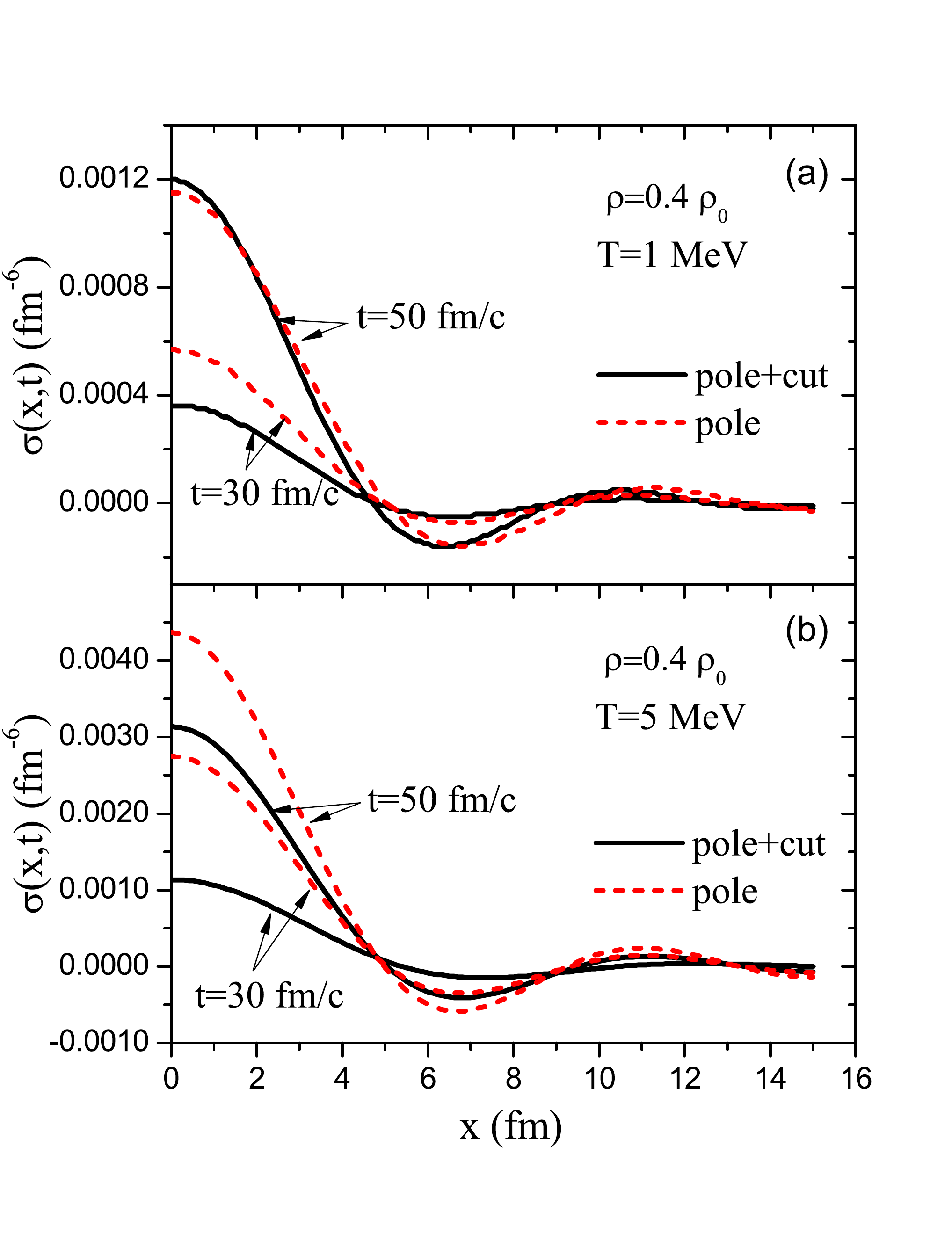}
\vspace{-1.5cm}
 \caption{\label{fig10}(Color online) Density correlation function as a function of distance between two space location $x=|\vec{r}-\vec{r}~'|$ for initial density$\rho =0.4\rho _{0} fm^{-3} $and temperature $T=1MeV$(a) and $T=5MeV$(b) at times $t=30,50fm/c$. Solid and dotted lines are complete calculations and pole contributions, respectively. }
\vspace{-0.0cm}
\end{figure}
In Fig. (\ref{Eq9}) calculations with cut-off are compared with the complete calculations of the density correlation function at temperatures $T=1MeV$and $T=5MeV$. Indeed, we see that the cut-off provides a good approximation at this initial density. At the initial density of $\rho =0.4\rho _{0} fm^{-3} $, there is no visible local minimum of the spectral intensity. In this case, we cut the integrations over the wave number at $k_{cut} =0.8fm^{-1} $. In Fig.(10), we compare the density correlation function calculated with the complete spectral intensity and with those calculated with pole contribution with cut-off at temperatures $T=1MeV$and $T=5MeV$. We observe that the results obtained with pole approximations are rather sensitive to the cut-off wave number.

\section{Conclusions}

 We investigate early growth of density fluctuations in nuclear matter in using the SMF approach. In the linear response framework, employing the method of one-sided Fourier transform, it is possible to carry out a nearly analytical treatment of the correlation function of density fluctuations. The density correlation function provides a very useful information about the early condensation mechanism during the liquid-gas phase transformation of the system. The one-sided Fourier transform method involves a contour integration in the complex frequency plane. In earlier investigations, this contour integral was evaluated by keeping only effects of the unstable collective poles, and the effects of non-collective poles were ignored. In the spectral intensity of the density correlation function, the collective pole contribution has an important drawback that it leads to a divergent behavior as the wave numbers approaches its upper limit, $k\to k{}_{\max } $. In previous studies to cure this problem, the integration over wave numbers was terminated at a cut-off wave number. Since short wavelength fluctuations cannot grow in time, cut-off provides a relatively good approximation. However, the complete description is always desirable. In this work, we consider symmetric nuclear matter in non-relativistic framework of the SMF approach.  We calculate the spectral intensity of the correlation function, including the effects of collective poles and non-collective poles in terms of the cut contribution in the complex frequency plane.  The cut contribution also has a divergent behavior with opposite sign, as wave numbers approach the upper limit, $k\to k{}_{\max}$.  As a result divergent behaviors of pole and cut contributions cancel each other out to produce a nice regular behavior of the spectral intensity as a function of wave number. This allows us to have a complete description of the correlation function of density fluctuations. It will be very interesting to carry out similar calculations of correlation functions for charge asymmetric nuclear matter in the semi-classical and the quantal framework by employing non-relativistic and relativistic versions of the SMF approach.

\begin{acknowledgments}

 S.A. gratefully acknowledges TUBITAK and the Middle East Technical University for partial support and warm hospitality extended to him during his visits. This work is supported in part by US DOE Grant No. DE-FG05-89ER40530, and in part by TUBITAK Grant No. 114F151.

\end{acknowledgments}

\end{document}